\documentclass[12pt,a4paper]{article}
\usepackage{graphicx,epsfig,epsf,rotating,fancyhdr,amsmath,natbib}
\usepackage{txfonts}

\def\etal{{et al.}}

\def\3o{O~{\sc ii}}
\def\4o{O~{\sc iv}}

\title{\bf \vspace*{-15mm}Sporadic Long-term Variability in Radio Activity from a Brown Dwarf }
\author{A. Antonova$^1$, J.G. Doyle$^1$, G. Hallinan$^2$, A. Golden$^2$ and C. Koen$^3$}
\date{$^1$Armagh Observatory, College Hill, Armagh,  N. Ireland \\ $^2$Computational
 Astrophysics Laboratory, National University of Ireland, Galway, Ireland \\ 
 $^3$Department of Statistics, University of the Western Cape, Bellville, South Africa}
\begin{document}
\maketitle

\abstract {Radio activity has been observed in a large variety of
stellar objects, including in the last few years, ultra-cool
dwarfs. To explore the extent of long-term radio activity in
ultra-cool dwarfs, we use data taken over an extended period of 9
hr from the Very Large Array of the source 2MASS J05233822-1403022
in September 2006, plus data taken in 2004. The observation taken
in September 2006 failed to detect any radio activity at 8.46 GHz. A
closer inspection of earlier data reveals that the source varied from
a null detection on 3 May 2004, to $\approx$95 $\mu$Jy on 17 May
2004, to 230 $\mu$Jy on 18 June 2004. The lack of detection in
September 2006 suggests at least a factor of ten flux variability at
8.46 GHz. Three short photometric runs did not reveal any optical
variability. In addition to the observed pulsing nature of the
radio flux from another ultra-cool source, the present observations suggests
that ultra-cool dwarfs may not just be pulsing but can also display long-term
sporadic variability in their levels of quiescent radio emission. The lack of 
optical photometric variability suggests an absence of large-scale spots at 
the time of the latest VLA observations, although small very high latitude 
spots combined with a low inclination could cause very low amplitude 
rotational modulation which may not be measurable. We discuss this large
variability in the radio emission within the context of both gyrosynchrotron 
emission and the electron-cyclotron maser, favoring the latter mechanism.}\ 
   

%

\section{Introduction}
For ultra-cool stars and brown dwarfs, both H$\alpha$ and X-ray emission
decline in strength so that few field objects later than L5 exhibit evidence of
activity at these frequencies (Schmidt \etal\ 2007 and references therein). This 
reduction in both chromospheric and coronal
emission is in broad agreement with some theoretical models, i.e. the cool,
dense atmospheres implying low ionization fractions and thus high electrical
resistivities, leading to a decoupling of magnetic lines from the upper
atmosphere (Mohanty \etal\ 2002).

Given the above considerations, it was generally assumed that radio
emission from cool dwarfs and brown dwarfs would be either weak or
absent. This was supported by the G\"udel \& Benz (1993) empirical
correlation between radio and X-ray emission which holds over many
orders of magnitude. However, the detection by Berger \etal\ (2001)
of radio emission from the M9 field brown dwarf LP 944-20 violated
the above relation by four orders of magnitude. Further observations
produced more radio active ultra-cool dwarfs (Berger 2002, Burgasser
\& Putman 2005, Berger 2006, Phan-Bao \etal\ 2007). In Fig.~1 we
plot the ratio of the radio to bolometric luminosity for all known
radio active ultra-cool dwarfs compared to a selection of
early-to-mid M dwarfs taken from the literature. Until recently, the
detected radio emission has been interpreted as incoherent
gyrosynchrotron radiation from a population of non-thermal electrons
(Berger 2002, Burgasser \& Putman 2005, Berger 2006, Osten \etal\
2006).

\begin{figure*}[ht!]
\vspace*{8.5cm}
\includegraphics{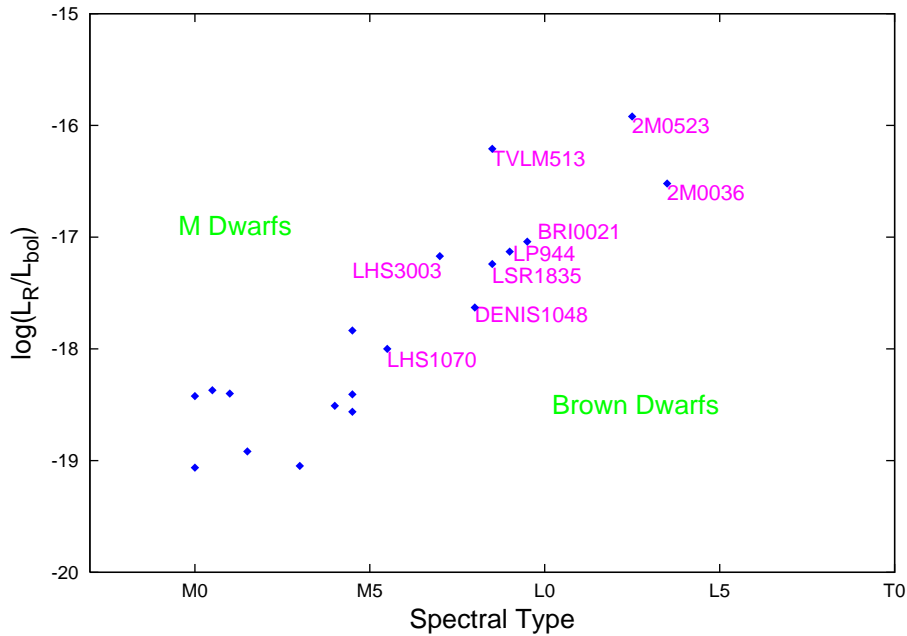}
\caption{Ratio of the radio to bolometric luminosity for all known
radio active ultra-cool dwarfs, compared to a selection of
early-to-mid M dwarfs. The present 2006 radio data for 2M0523 is at
least a factor of ten lower.}
\label{Fig1}
\end{figure*}

Berger \etal\ (2005) conducted simultaneous 4.9 GHz and 8.5 GHz observations of the 
L3.5 brown dwarf 2MASS J00361617+1821104 which yielded highly polarized emission in both  
bands. Furthermore, the emission was both periodic and variable, with the average 
flux and significance of the variability higher at 4.9 GHz. Due to the low net 
circular polarization
and the broadband persistent nature of the emission, Berger \etal\ attributed it to the  
gyrosynchrotron process. Hallinan \etal\ (2006) detected periodic emission from the M9
dwarf TVLM 513-46546 with the similar properties to those detected from 2MASS J00361617+1821104
but, using geometrical constraints, suggested an alternative 
process, the electron cyclotron maser instability as the source of the
dominant radio emission. In subsequent observations of the 
same dwarf, Hallinan \etal\ (2007) reported on 100 \%  circularly polarized bursts
with a periodicity that was in very good agreement with the rotational period of the 
dwarf, as confirmed by I-band photometric observations. From the duration of the 
bursts they constrained the size of the emitting source region to be less than 0.22 
times the radius of the dwarf, estimating the brightness temperature of the radio 
emission. Those bursts conclusively confirmed that at least part of the emission was 
due to the electron cyclotron maser instability. 

Considering the similarities in the properties of the radio emission of these two 
dwarfs, it is possible that the same mechanism is responsible for the emission 
in both cases. The above observations suggest the probable existence of large-scale, 
stable magnetic fields on at least some ultra-cool dwarfs. In this Letter we report on 
high sensitivity VLA observations of the radio active L2.5 brown dwarf 2MASS 
J05233822-1403022, undertaken to investigate the production mechanism of the radio 
emission, as well as to search for rotational modulation of the emission.

\vspace*{-3mm}
\section{Observations and Data Reduction}
\subsection{Radio}
The observations of 2MASS J05233822-1403022 (hereafter 2M0523) were
conducted with the NRAO Very Large Array (VLA)\footnote{The National Radio
Astronomy Observatory is a facility of the National Science Foundation operated
under cooperative agreement by Associated Universities, Inc.} for a duration of
$\approx$ 9 hours as part of a monitoring program on several previously
detected ultra-cool dwarfs. 2M0523 was observed on 2006 September 23, at
8.45 GHz (3.6 cm). During the observations the array consisted of 23 antennas
in the B configuration. We used the standard continuum mode with 2 $\times$ 50
contiguous bands, sampling every 10 s. The flux density calibrator was
3C138, while the phase calibrator was 0513-219 with the time on source in a
single scan being 12 minutes, before moving to the phase calibrator for 2
minutes.

For comparison, we obtained a set of observations of the same source from the
public archive of the NRAO Data Archive System. These observations were
conducted on 2004 May 3, May 17 and June 18, at 8.45 GHz, each for a duration
of $\sim$2 hours, with array configurations C, DnC and D respectively,
in standard continuum mode with 2 $\times$ 50 contiguous bands. The same
calibrators (3C48 \& 0513-219) were used.

Data reduction was carried out with the Astronomical Image Processing System
(AIPS) software package. The visibility data was inspected for quality both
before and after the standard calibration procedures, and noisy points were
removed. For imaging the data we used the task IMAGR. We also CLEANed the
region around each source and used the UVSUB routine to subtract the resulting
source models for the background sources from the visibility data. For
examining the light curves we used the AIPS task DFTPL.

\vspace*{-3mm}
\subsection{Optical}

The optical observations were made in the $I_C$ band, with a CCD
camera attached to the 1.9~m telescope of the South African
Astronomical Observatory. The 1024$\times$1024 CCD SITe chip gives a
camera field of view of 25 arcmin$^2$ on the telescope, which
allowed measurement of four bright comparison stars. The camera was
operated in $2\times2$ prebinning mode throughout. Calibration
flat-fields were obtained during evening twilight, under perfect
photometric conditions.

The photometric reductions were performed with a modified version
of DOPHOT (Schechter \etal\ 1993). The profile-fitted
magnitudes were considerably more accurate than the aperture
magnitudes, and only the former were retained.

An observing log is given in Table 1:
the first two sets of measurements were published in Koen (2005),
where time series plots of those data can be seen.

\begin{table}
\caption{A summary of $I_C$ band observations of 2MASS J05233822 -1403022 in December 
2004 and 2006. The penultimate column gives the r.ms. scatter in the data,
and the last column the range of r.m.s. scatter for the five brightest stars
in the field of view.}

\begin{center}
\begin{tabular}{ccccccccc}
Start time & Run & Exp & $N$ & $\sigma$ & $\sigma$\\
(HJD~2450000+)& (hours) & (s) & & (mmag) & (mmag)\\
 & & & & \\
 3363.3115 & 1.6 & 120& 42 & 7 & 6-12\\
 3366.2931 & 3.0 & 90& 96 & 7 & 6-9 \\
 4088.4897 & 1.7 & 100 & 23 & 5 & 4-8
\end{tabular}
\end{center}
\end{table}

\begin{figure*}[hbt!]
\vspace{8.0cm}
\includegraphics{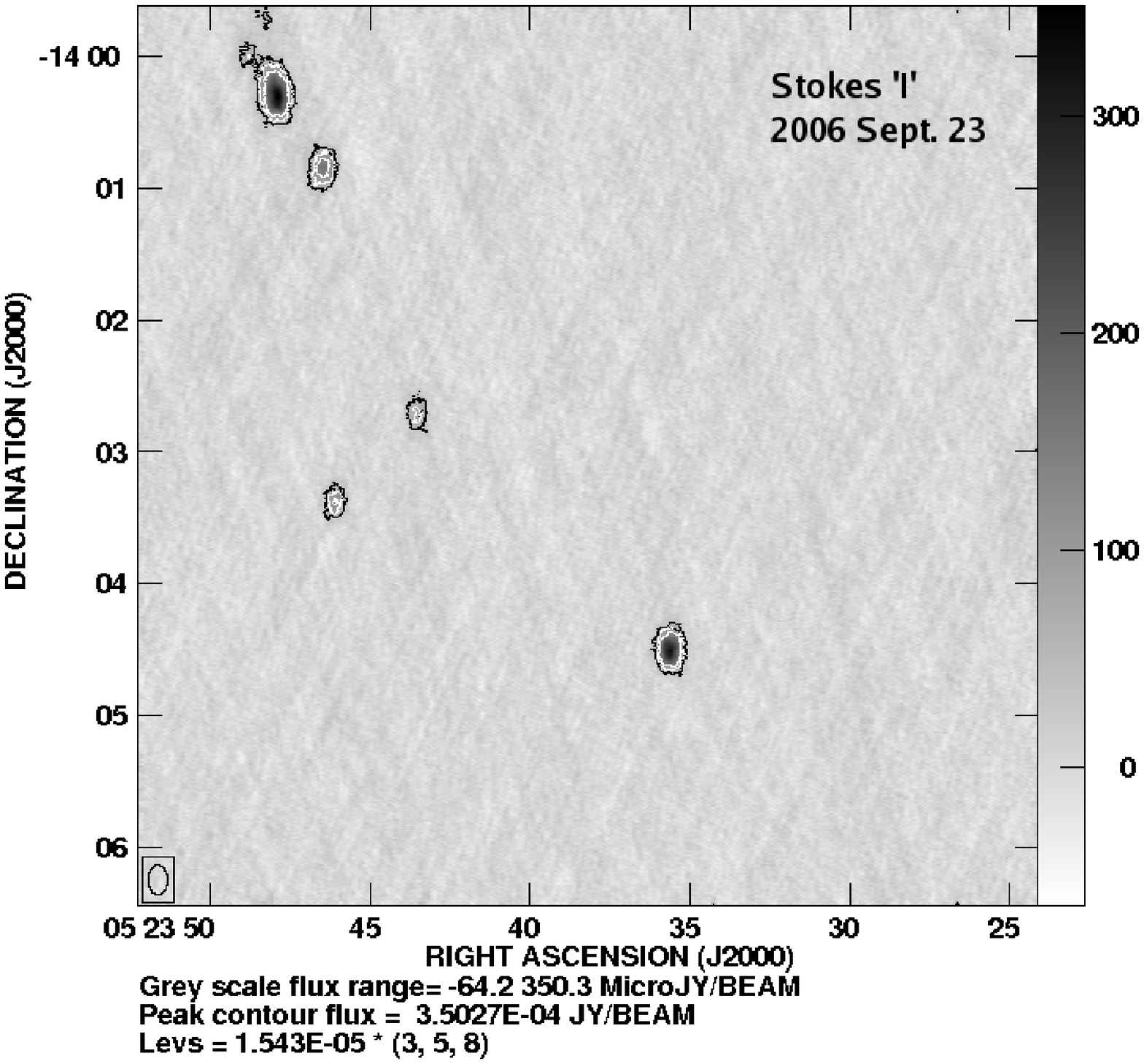}
\includegraphics{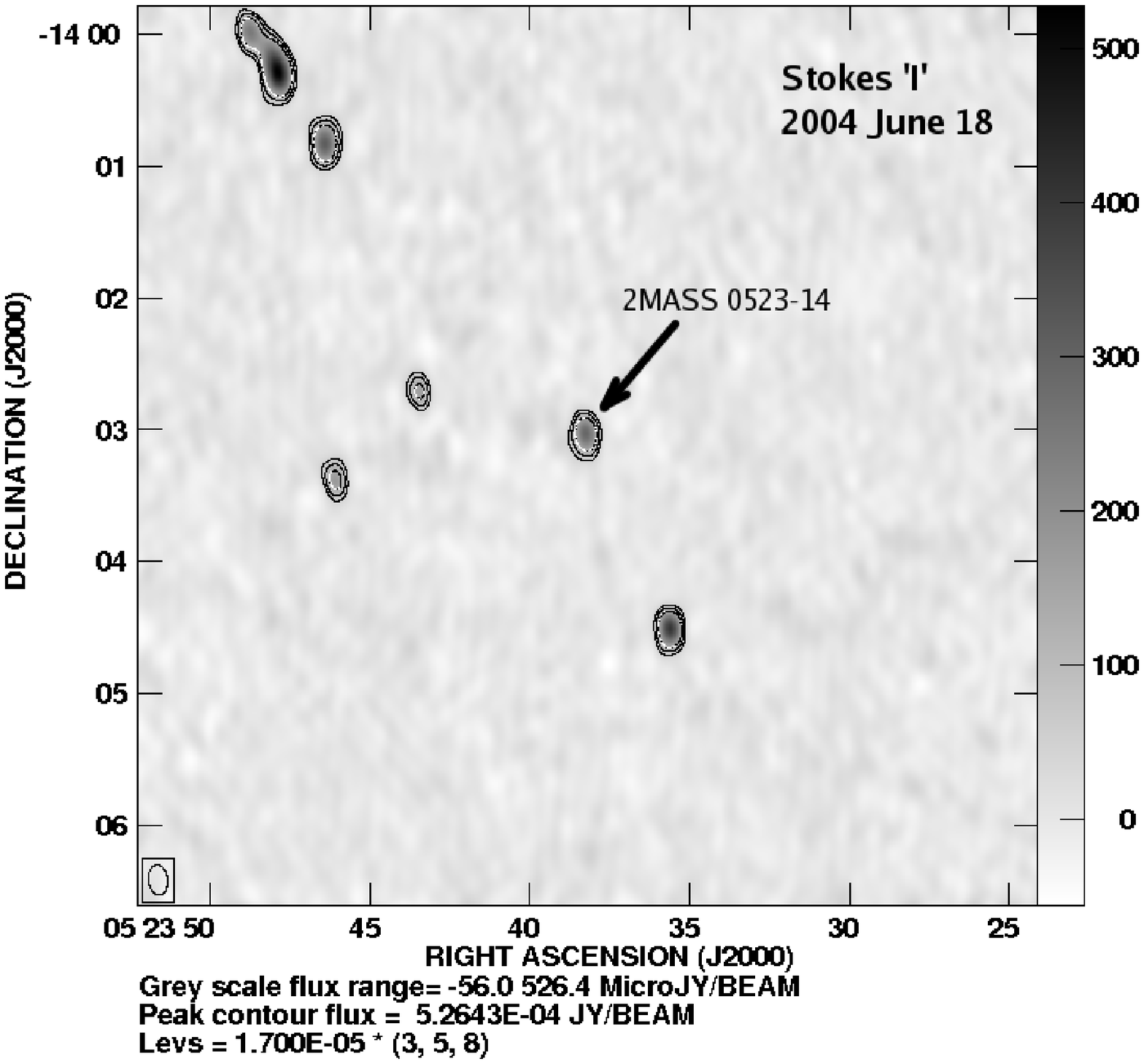}

\vspace*{-0.5cm}
\caption{(left) VLA map of the 2MASS J05233822-1403022 field at 8.46 GHz in total 
intensity (Stokes I), on 23 September 2006 and (left) on 18 June 2004 (right). The 2006
data was taken in the B configuration while the June 2004 data was in the D
configuration.}
\label{Fig2}

\end{figure*}
\vspace*{-3mm}
\section{Results}
2M0523, first discovered by Cruz \etal\ (2003) is a L2.5 brown dwarf
located at a distance of 13.4 pc. It was previously reported as
radio active by Berger (2006), with a flux of 231 $\mu$Jy (see
Fig.~1). Thus we expected that imaging the visibilities of our data
would reveal a source at the position of the dwarf. To our surprise,
no source was present at that position. Fig. 2a shows the image of
the  field in total intensity (Stokes I), with the background
sources present. The r.m.s. of the map is $\approx$15 $\mu$Jy, which
gives a 3 $\sigma$ upper limit of 45 $\mu$Jy.

Our re-analysis of the 2004 archive data sets revealed an
interesting picture. There is no detection on May 3, with a 3
$\sigma$ upper limit of 42 $\mu$Jy. We did detect the source on May
17, with a flux of 95 $\mu$Jy. Imaging the June 18 data revealed a
yet stronger source with a flux of 230 $\mu$Jy (Fig. 2b). The r.m.s
values for the May 17 and June 18 maps are 19
$\mu$Jy and 17 $\mu$Jy respectively.

Due to differences in the setup, the resolution of the datasets was
different. In Fig 2a, the lower resolution of the image was done by: 1)
retrieving the dimensions of the CLEAN restoring beam from the header of the
low resolution image (i.e. the 2004 data) and 2) specifying those dimensions
in the task IMAGR when imaging the high resolution image. The same
background sources are clearly visible in both maps with the absence of a
source at, or even near the position of 2M0523 in the 2006 September 23 data.
In order to check whether the 2004 May 17 or June 18 detection is dominated by
a large flare, we also looked at the total intensity light-curves versus time.
No flaring activity was detected.

In the optical, the present object was monitored for variability in
December 2004 and 2006 (see Table~1). Although there is no evidence
for optical brightness variations on timescales shorter than 2-3
hours, the question of changes on timescales of days remains open.
In order to investigate this, we use measurements for the five
brightest stars ($16.4 \le I_C \le 17.1$, assuming $I_C=16.6$ for
2M0523 -- see Koen 2005). Differential magnitudes of each
object were determined with respect to the means of the five stars,
for each time-point observation. This gave the r.m.s. values quoted in
the Table. The formal standard errors in the nightly mean magnitudes
for the stars are very small -- the largest is about 3.5 mmag (see
Table 1). This does not, of course, take account of systematic
errors, which unfortunately dominate. The night-to-night changes for the
target object are larger, but only marginally. We conclude that there is
currently no evidence for optical variability in 2M0523.

\vspace*{-3mm}
\section{Discussion}
For earlier type M dwarfs, both the quiescent and flaring coronal radio
emission are generally attributed to gyrosynchrotron emission from a
non-thermal population of mildly relativistic electrons with a power-law
distribution. Such emission is associated with low levels of
polarization and implies magnetic field strengths of
up to several hundred Gauss, electron densities of $\lesssim 2 \times 10^{9}$
cm$^{-3}$ and brightness temperatures of 10$^{8}$ -- 10$^{9}$ K.

There are however some mid- M dwarfs whose radio emission
characteristics are not well explained by the gyrosynchrotron
mechanism. UV Ceti, a dM5.5e star, is one of these and has been
observed to emit bright, 100 \% polarizied flares (Kellett \etal\ 2002). 
Similarly, AD Leo, a dM3.5e star, has shown a series of
high, frequency highly polarized radio pulses during a flare (Bastian
\etal\ 1990). Also, Burgasser \& Putman (2005), although noting that
the quiescent radio emission from two ultra-cool dwarfs was
consistent with optically thin gyrosynchrotron emission, suggested
that two detected flares implied a coherent emission process. These
had high brightness temperature ($10^{13}$~K), were of short
duration ($\approx$4--5 min) and were highly circular polarizied 
($\approx$100\%).

The characteristics of many of these bursts are consistent with the electron 
cyclotron maser process. For the maser operation, a
population inversion in the electron distribution is needed, as well
as a relatively strong magnetic field and low-density plasma, so
that the electron-cyclotron frequency $\nu{_c}$ is greater than the
plasma frequency $\nu{_p}$.

As to the instability at the heart of the process, different
distributions has been proposed, e.g. Wu \& Lee (1979) suggested a
loss-cone for the terrestrial auroral kilometric emission while
Pritchett (1984) suggested a shell distribution. It has since been shown that the
shell instability is much more efficient than the loss-cone instability as 
a source of AKR (see Treumann 2006 for a review). It is
suggested that the main source for any strong electron-cyclotron
maser is found in the presence of a magnetic-field-aligned electric
potential drop which has several effects. For example, it can dilute
the local plasma to such an extent that the plasma enters the regime
in which the electron-cyclotron maser becomes effective and favors
emission in a direction roughly perpendicular to the ambient
magnetic field. This emission is the most intense, since it implies
the coherent resonant contribution of a maximum number of electrons
in the distribution function. What is more, such an instability can be sustained
over a range of heights above the stellar surface, thus producing pseudo
broadband, coherent radio emission, which would explain the simultaneous
detection of both 8.4 and 4.9 GHz emission from TVLM 513-46546 (Hallinan 
\etal\ 2006, 2007) and 2MASS0036+18 (Berger \etal\ 2002, 2005). This would imply  
kilogauss fields on ultra-cool dwarfs, which has been recently confirmed by 
Reiners \& Basri (2007). 

Whether such a model applies in this instance is unclear since we do not have
pulsed emission. There is no evidence which can conclusively distinguish between the
maser and the gyrosynchrotron processes as the source of the `quiescent'  
radio emission detected from any UCD thus far. Assuming the persistent emission is
isotropic, gyrosynchrotron radiation from an extended corona (as suggested for
quiescent radio emission from early dMe stars) would imply magnetic fields up to a few 
hundred Gauss, brightness temperatures limited to a few times $10^{9}$ K and a source region
of $\sim$ 2 - 3 times the radius of the dwarf. The presence of such emission requires a
constant supply of electrons accelerated to coronal energies. The most commonly 
prescribed mechanism for such acceleration is magnetic reconnection leading to micro-flaring.
Yet in the cool, dense atmospheres of UCDs the magnetic field is largely decoupled from
the atmosphere (Mohanty \etal\ 2002), and as a result the generation and propagation of
magnetic stresses which can lead to magnetic reconnection is inhibited at the dwarf's
surface. 

However, if we assume that gyrosynchrotron is possible and since it is roughly proportional 
to the injection rate of electrons (G\"udel 2002), thus if by some process the injection 
rate is reduced (for example a reduced level of micro-flaring), this would mean a reduced level of radio 
emission. In the case of 2MASS 0523 there was no evidence for flare activity during 
the enhanced flux levels detected in June 2004, suggesting instead a steady rise in 
the overall flux levels over a period of at least five weeks. Nevertheless, the 
`quiescent' emission from this source displays a ten-fold variation in the flux levels, 
which is similar to what is observed for the pulsed coherent emission from TVLM 513 
(Hallinan \etal\ 2007). Therefore an alternative explanation for this large scale 
variability may be given in terms of a large dipolar-field and the electron-cyclotron 
maser (such emission would require  
magnetic field strengths of the order of several kG, brightness temperatures 
$\gtrsim 10^{11}$ K and source region of up to a few tenths of the stellar radius).
For such a field configuration, the plasma density would not be homogeneous throughout 
the magnetosphere. At present neither mechanism can be ruled out, yet it is notable that 
the large variations in flux with time are similar to what is observed for the pulsed 
emission from UCDs. 

Hallinan \etal\ (2006) showed that TVLM 513-46546 has a well defined spot
structure. The present I-band data suggest a lack of large-scale spots on
2M0523 in December '04 and '06, although whether spots were present
during the VLA observations of June '04 in unclear. The possible lack of
spots detected from 2M0253 can be explained by a series of factors. For
example, 2M0253 is of later spectral type and hence the spots would have a
lower contrast to the photosphere therefore reducing the chances of detection.
It should also be noted that very high latitude spots combined with a low
inclination could cause very low amplitude rotational modulation which may
not be measurable. Furthermore, truly polar spots do not cause rotational
modulation of the stellar brightness at all.

Such spot activity and hence the presence of increased magnetic reconnection
in the lower atmosphere could lead to higher plasma levels in the outer atmosphere.
Such activity could disrupt the stable electric fields in the magnetosphere
of the ultra-cool dwarf. While the plasma
population in the magnetosphere may re-stabilize after such an
injection of plasma, the maser emission would probably switch off.
The present observations therefore suggests that ultra-cool dwarfs
may not just be pulsing but may also be sporadic in their emission levels,
which leads to the question of how many of these ultra-cool dwarfs
are radio active. For example, Berger (2006) suggested $\approx$10\%.
However, the results presented here show that this value is dependent on the 
assumption that these objects are stable emitting sources. Long-term  
multi-frequency monitoring of the behaviour of a  small 
sample of these objects should be undertaken before further large-scale
surveys commence, in addition to further simultaneous optical and radio
monitoring in order to assert the nature of activity in these objects.

However, the above interpretation of the `quiescent' radio emission (although one
may question the use of the term quiescent) comes with the obvious health warning that 
more data is needed before we would be in a position to identify with certainty the 
mechanism of the `quiescent' radio emission in UCDs. \\

{\bf Acknowledgements}
Armagh Observatory is grant-aided by the N. Ireland Dept. of Culture, Arts \&
Leisure. The authors thank Stephen Bourke for his help with IRAF and Kathlin
Olah for advice on spot modelling. We also gratefully
acknowledge the support of the HEA funded Cosmogrid project and Enterprise
Ireland under the grant award SC/2001/0322, plus PPARC for a visitor grant.\\

\end{document}